\documentclass[twocolumn,amssymb,amsmath,floats,showpacs,pre]{revtex4-1}

\usepackage{amsmath,amssymb,bm}
\usepackage{graphicx}
\usepackage{xcolor}
\usepackage{multirow}

\begin{document}

\title{Generalization of core percolation on complex networks
}

\author{N. Azimi-Tafreshi}

\affiliation{Physics Department, Institute for Advanced Studies in Basic Sciences, 45195-1159 Zanjan, Iran }

\author{S. Osat}

\affiliation{Quantum Complexity Science Initiative, Skolkovo Institute of Science and Technology, Skoltech Building 3, Moscow, 143026, Russia}

\author{S.~N. Dorogovtsev}
\affiliation{Departamento de F{\'\i}sica da Universidade de Aveiro $\&$ I3N, Campus Universit\'ario de Santiago, 3810-193 Aveiro, Portugal}
\affiliation{A.F. Ioffe Physico-Technical Institute, 194021 St. Petersburg, Russia}


\begin{abstract}
We introduce a $k$-leaf removal algorithm as a generalization of the so-called leaf removal algorithm. In this pruning algorithm, vertices of degree smaller than $k$, together with their first nearest neighbors and all incident edges are progressively removed from a random network. As the result of this pruning the network is reduced to a subgraph which we call the Generalized $k$-core ($Gk$-core). Performing this pruning for the sequence of natural numbers $k$, we decompose the network into a hierarchy of progressively nested $Gk$-cores. We present an analytical framework for description of $Gk$-core percolation for undirected uncorrelated networks with arbitrary degree distributions (configuration model). To confirm our results, we also derive rate equations for the $k$-leaf removal algorithm which enable us to obtain the structural characteristics of the $Gk$-cores in another way. Also we apply our algorithm to a number of real-world networks and perform the $Gk$-core decomposition for them.
 \end{abstract}


\maketitle

\section{Introduction}
Structural decomposition of complex networks providing classification the vertices into different subsets is one of effective approaches for studying the structural properties of networks. As a primary and well-known example, one can indicate the $k$-core decomposition, which is an efficient technique for uncovering structural properties of large networks \cite{Alvarez, Carmi}. The $k$-core of a network is defined as the largest subgraph whose vertices have degree at least $k$ \cite{Seidman}. There is a pruning algorithm enabling one to obtain $k$-core subgraphs for a given network: at each step, a vertex of degree less than $k$ is randomly chosen and removed. The pruning is continued until no further removal is possible. As the result of this pruning the network is decomposed to a set of enclosed $k$-cores. The vertices belonging to higher (more central) cores are more strongly connected. It was also shown that the vertices of inner core are more influential spreaders in epidemic processes \cite{Kitsak}. A giant $k$-core emerges above a percolation threshold \cite{k-corepercolation}. The most remarkable result is that for $k\geq 3$ the giant $k$-core shows a discontinues hybrid phase transition combining discontinuity and a critical singularity \cite{k-corepercolation, critical}. Furthermore, generalized models for $k$-core percolation have been studied on interdependent and multiplex networks, which reveal more features than the ordinary $k$-core percolation problem on single networks \cite{generalk-core1, generalk-core2}.

Another key subgraph of a random network is simply called its core. These subgraphs significantly differ from the $k$-cores. A core of an undirected network is obtained only by a pruning algorithm in contrast to the $k$-core which is, in addition, defined by a specific constraint on the connectivity of its vertices. The pruning algorithm producing a core is called the leaf removal algorithm. It was introduced by Karp and Sipser \cite{Karp}. In this pruning algorithm, a vertex of degree one (a leaf) is randomly chosen and removed together with its neighbor and all incident edges. The algorithm is continued until no leaves remain. The resulting subgraph is formed by some isolated subgraphs and the giant one which is called the core. For the Erd\H{o}s--R\'enyi (ER) random graphs, Bauer and Golinelli showed that the core percolation threshold is located at the mean degree $\langle q\rangle = e=2.718...$, so that above this point the network contains the giant core while below the threshold the size of the giant core is zero \cite{Bauer}. The core structure and the phase transition at $\langle q\rangle = e$ is related to a number of phenomena in physics such as conductor-insulator transitions \cite{local} and replica symmetry breaking in minimal vertex cover \cite{replica}. Moreover it was shown that the formation of the core is related to controllability robustness \cite{control1,control2}
and some combinatorial optimization problems such as maximum matching and minimum vertex cover \cite{Karp, replica,Weight:whbook2005}. Also a generalized leaf removal process, which is applicable in minimum dominating set problem, has been introduced in \cite{MDS}. Using a time-dependent analysis, people have studied the core percolation related to this generalized leaf-removal algorithm.

In this paper, we generalize definition of the leaf to the ``$k$-leaf'', defined as a vertex of degree less than $k$.
In this algorithm we remove recursively a $k$-leaf together with all its first neighbors and their
incident edges. Following this pruning algorithm, the network is decomposed to a hierarchy of nested cores, similarly to the ordinary $k$-core decomposition. We call this structure the Generalized $k$-core ($Gk$-core). In this notation, the ordinary core is represented by the $G2$-core. The vertices belonging to inner $Gk$-cores and also their first neighbors are of high degree and well connected. Analytical calculation is possible only for the networks with locally tree-like structure. For these kind of networks, and using the generating function technique, we study the structural transitions and emergence points of the $Gk$-core subgraphs.

The $k$-leaf removal algorithm can be also considered as the inducing effect, introduced by Zhao et al. \cite{Zhao}. In the inducing process, a collapsed vertex $i$ will induce its remaining neighbors, to be collapsed if vertex $i$ has less than $k$ remaining neighbors. In \cite{Zhao}, the inducing effect together with the spontaneous collapsing process, leads to the emergence of other subgraphs, called protected cores.

The leaf removal algorithm is a Markovian process. We describe evolution of the network structure
during the pruning process by applying rate equations. The rate equations have been derived for the ordinary leaf removal algorithm on undirected and directed graphs \cite{Weight:whbook2005,Weigt,coredir}. This approach provided the size and the emergence point of the ordinary core.
In this paper, we also derive rate equations for the degree distribution of a network during the execution of the $k$-leaf algorithm, which enables us to obtain the structure of the $Gk$-cores in an alternative way.

The paper is organized as follows. In Sec.~\ref{analytic}, we present an analytical framework to study $Gk$-core percolation for random networks with arbitrary degree distributions. We apply our general results to the
Erd\H{o}s--R\'enyi and scale-free networks. We compare our results with numerical simulations. In Sec.~\ref{rate}, we derive the rate equations for the $k$-leaf removal algorithm and using these equations we find in another way how $Gk$-cores are organized.
In Sec.~\ref{real}, a set of real-world networks are analyzed in the framework of our approach.
\begin{figure}
\begin{center}
\scalebox{0.5}{\includegraphics[angle=0]{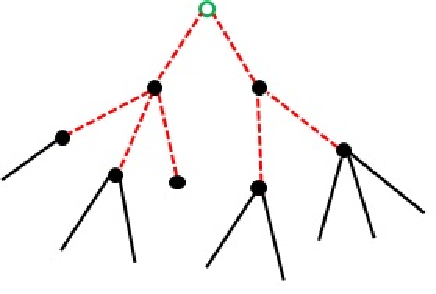}}~~~~
\end{center}
\caption{(color online). The open green dot shows a $k$-leaf. Once a $k$-leaf is selected, the $k$-leaf together with the dashed red edges are removed.
}
\label{f11}
\end{figure}
\section{Analytical framework}
\label{analytic}
Let us consider an uncorrelated network with an arbitrary degree distribution $P(k)$. To produce a generalization of the core subgraph, we use the following pruning algorithm: at each step we randomly choose a $k$-leaf (i.e. a vertex of degree less than $k$), remove it together with its neighbors and all incident edges to the neighbors. Figure~\ref{f11} shows a $k$-leaf (open green dot) and the $k$-leaf removal process. As a result of the pruning, the degrees of some vertices change. The procedure is iterated until no vertices of degree less than $k$ remain in the network. The residual network, if it exists, is called $Gk$-core.

To find the size of the $Gk$-core, we classify the vertices into three groups: 1) $\alpha$-removable: the vertices that can become a leaf. 2) $\beta$-removable: the vertices that can become a neighbor of a leaf. 3) the vertices that are neither $\alpha$-removable nor $\beta$-removable and hence belong to $Gk$-core. Using the assumption that the network has a locally tree-like structure, we can write self-consistency equations for probabilities that
a random neighbor of a random vertex is
$\alpha$-removable, $\beta$-removable or a non-removable vertex. We call these probabilities $\alpha$, $\beta$ and $1-\alpha-\beta$, respectively. These probabilities are represented graphically in Fig.~\ref{f1}. Note that definition of these probabilities is the same as that already defined in \cite{corepercolation}. The difference is in the definition of the leaves.

At least one of the neighbors of a $\beta$-removable vertex must be $\alpha$-removable. Furthermore, an end vertex of a randomly chosen edge belongs to the $Gk$-core, if it has at least $k-1$ neighbors which belong to the $Gk$-core and none of its neighbors are of type $\alpha$. Taking into account these facts, we write the following two self-consistent equations:
\begin{figure}
\begin{center}
\scalebox{0.55}{\includegraphics[angle=0]{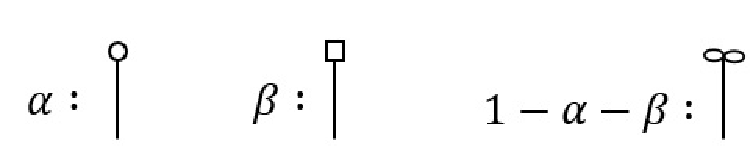}}~~~~
\end{center}
\caption{(color online). Schematic representation of the probabilities $\alpha$ and $\beta$.
}
\label{f1}
\end{figure}
\begin{figure}
\begin{center}
\scalebox{0.75}{\includegraphics[angle=0]{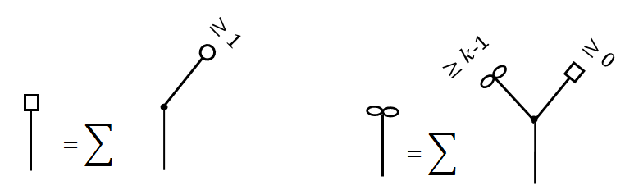}}~~~~
\end{center}
\caption{(color online) Graphical representation of the self-consistency equations for the probabilities $\beta$ and $1-\alpha-\beta$.
}
\label{f2}
\end{figure}
\begin{eqnarray}
1-\alpha-\beta&=&\sum_{q}\frac{q P(q)}{\langle q\rangle}
\nonumber\\
&&\times\sum_{s=k-1}^{q-1}
\left(
                      \begin{array}{c}
                        q-1 \\
                        s \\
                      \end{array}
                    \right)
(1-\alpha-\beta)^{s}\beta^{q-1-s},~~~
\label{eq1}
\!\!\!\!\!
\nonumber
\\[5pt]
\beta&=&1-\sum_{q}\frac{q P(q)}{\langle q\rangle}(1-\alpha)^{q-1}.
\label{eq2}
\!\!\!\!\!
\end{eqnarray}
The first equation represents the probability that an end vertex of a randomly chosen edge belongs to the $Gk$-core. $qP(q)/\langle q\rangle$ is the probability that the end vertex of a uniformly randomly chosen edge has degree $q$ and the combinatorial multiplier
$\left(
                      \begin{array}{c}
                        m \\
                        n \\
                      \end{array}
                    \right)$
gives the number of ways one can choose $n$ edges from a sample of $m$ edges. At least $k-1$ edges of $q-1$ edges (other edges than the starting one) must lead to the $Gk$-core. Eq.~(\ref{eq1}) also shows the probability that an end vertex of a randomly chosen edge is $\beta$-removable. At least one of the neighbors of a $\beta$-removable vertex must be a leaf, i.e. an $\alpha$-removable vertex. These two equation are schematically represented in Fig.~\ref{f2}.

\begin{figure}
\begin{center}
\scalebox{0.8}{\includegraphics[angle=0]{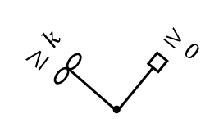}}~~~~
\end{center}
\caption{(color online). Schematic representation of the probability that a vertex belongs to the $Gk$-core, which is the relative size $n_{kc}$ of the $Gk$-core.
}
\label{f3}
\end{figure}

From Eq.~(\ref{eq1}), one can derive the following self-consistency equation for $\alpha$:
\begin{eqnarray}
\alpha=\sum_{q}\frac{q P(q)}{\langle q\rangle}
\sum_{s=0}^{k-2}
\left(
                      \begin{array}{c}
                        q-1 \\
                        s \\
                      \end{array}
                    \right)
(1-\alpha-\beta)^{s}\beta^{q-1-s}.
\label{eq3}
\!\!\!\!\!
\end{eqnarray}
The probabilities $\alpha$ and $\beta$ enable us to obtain the probability $n_{kc}$ that a randomly chosen vertex belongs to the $Gk$-core, which is also the relative size of the $Gk$-core. Fig.~\ref{f3} shows a schematic representation of this probability. A vertex is in the $Gk$-core if the vertex has at least $k$ neighbors which belong to the $Gk$-core. Hence we can write the following equation for the relative size of the $Gk$-core:
\begin{eqnarray}
n_{kc}=\sum_{q>k}P(q)
\sum_{s=k}^{q}
\left(
                      \begin{array}{c}
                        q \\
                        s \\
                      \end{array}
                    \right)
(1-\alpha-\beta)^{s}\beta^{q-s}.~~~~~~
\label{eq4}
\!\!\!\!\!
\end{eqnarray}
To be able to solve Eqs.~(\ref{eq1})--(\ref{eq4}) analytically, we rewrite these equations using generating functions \cite{generating}. For a network with a given degree distribution $P(q)$, the generating function $G(x)$ is defined as
\begin{equation}
G
(x)\equiv \sum_{q} P(q)x^{q}
\label{eq6}
.
\end{equation}
Hence, we obtain the following equations for $\alpha$, $\beta$ and $n_{kc}$ in terms of the generating function:
\begin{eqnarray}
\alpha&=&\frac{1}{\langle q\rangle}\sum_{s=0}^{k-2} \frac{(1-\alpha-\beta)^s}{s!} ~G^{(s+1)}(\beta),
\nonumber
\\[5pt]
\beta&=&1-\frac{G^{(1)}(1-\alpha)}{\langle q\rangle},
\nonumber
\\[5pt]
n_{kc}&=&G(1-\alpha)-\sum_{s=0}^{k-1} \frac{(1-\alpha-\beta)^s}{s!} ~G^{(s)}(\beta),
\label{eq5}
\end{eqnarray}
where we used the notation $G^{(s)}(x)$ for the $s$-th derivatives of $G(x)$.
\begin{figure*}[t]
\begin{center}
\scalebox{0.32}{\includegraphics[angle=0]{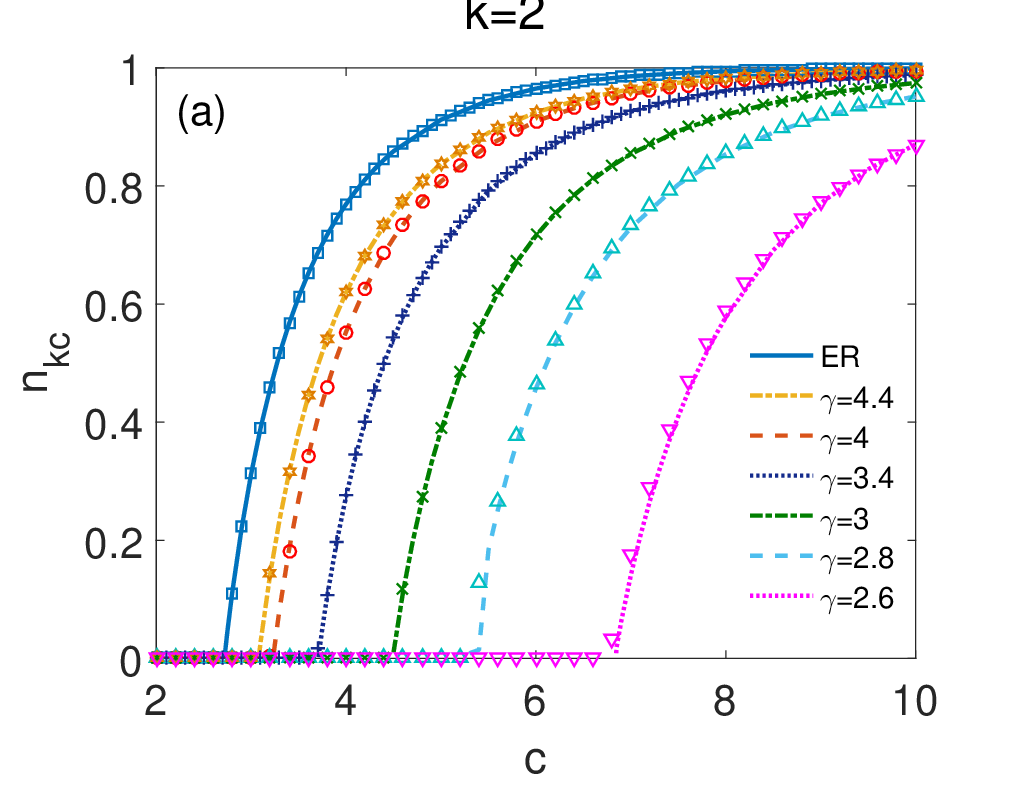}}
\scalebox{0.32}{\includegraphics[angle=0]{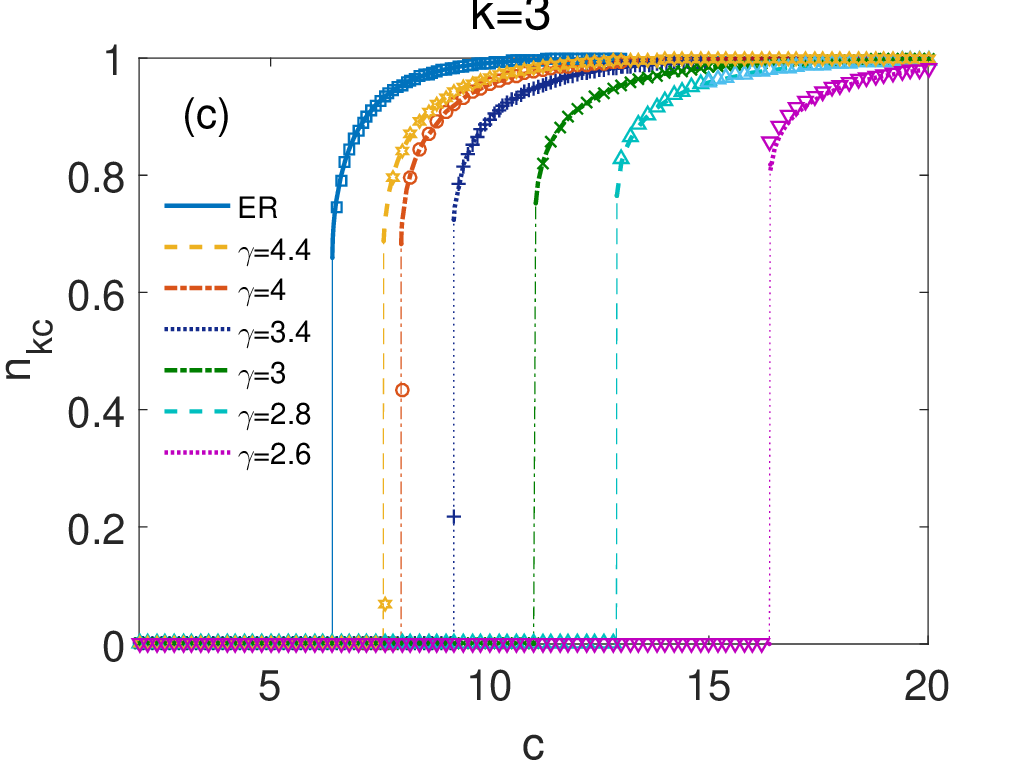}}
\scalebox{0.32}{\includegraphics[angle=0]{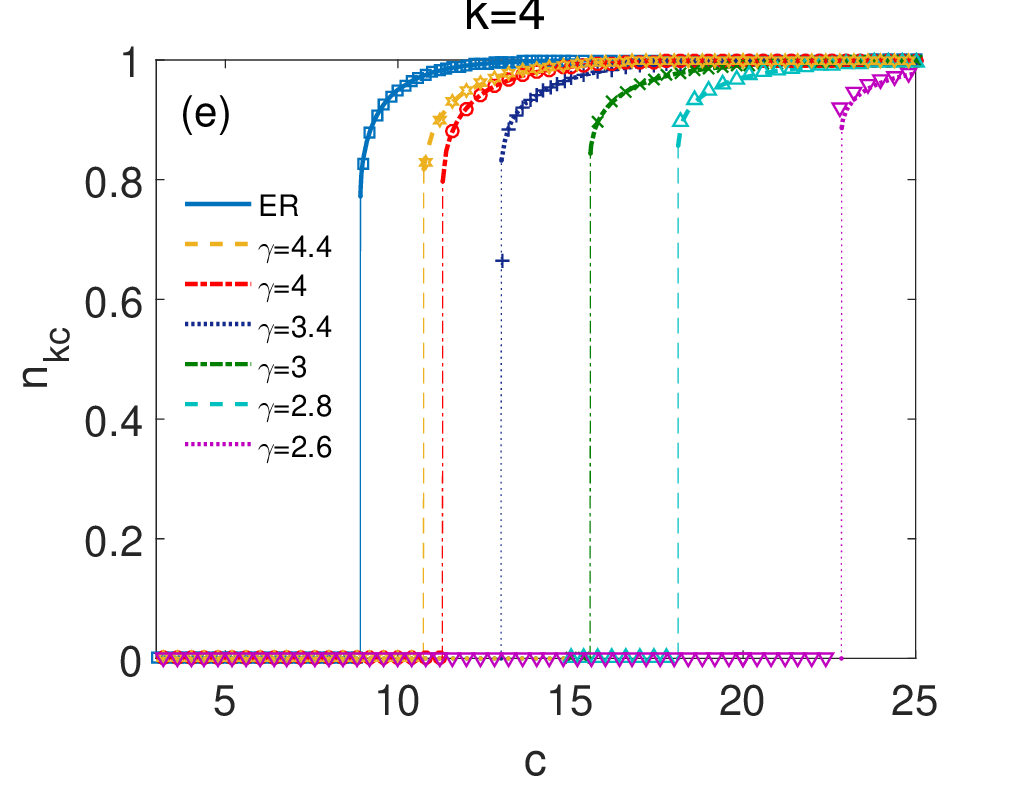}}
\\
\scalebox{0.32}{\includegraphics[angle=0]{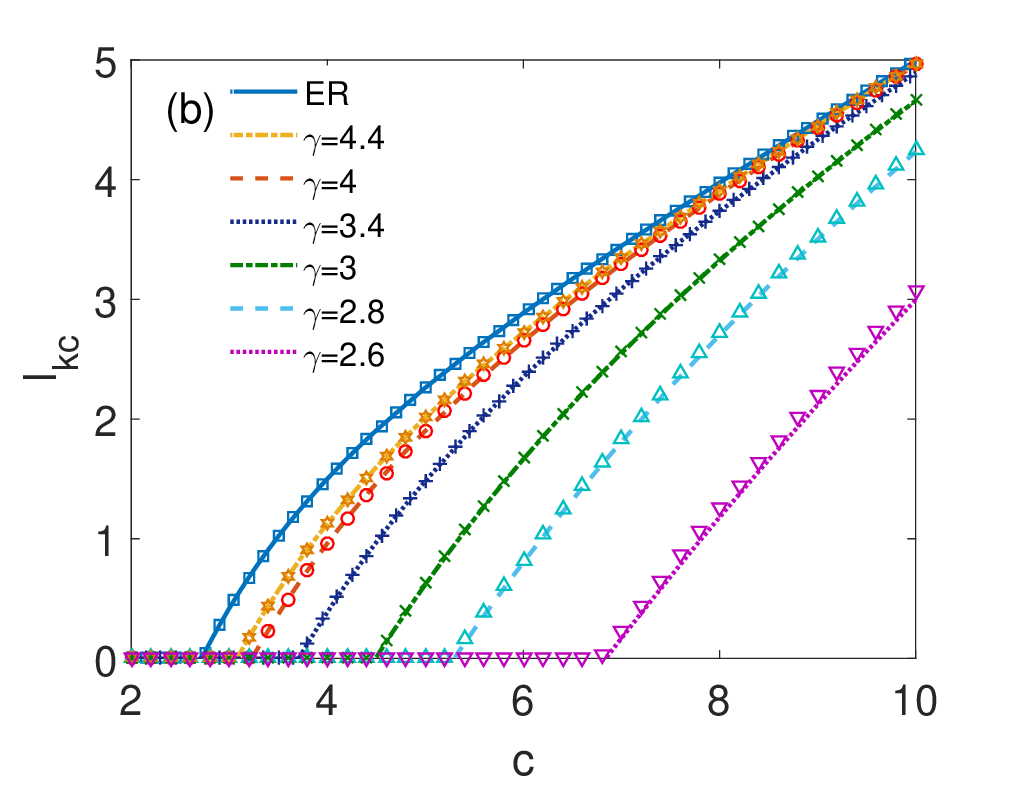}}
\scalebox{0.32}{\includegraphics[angle=0]{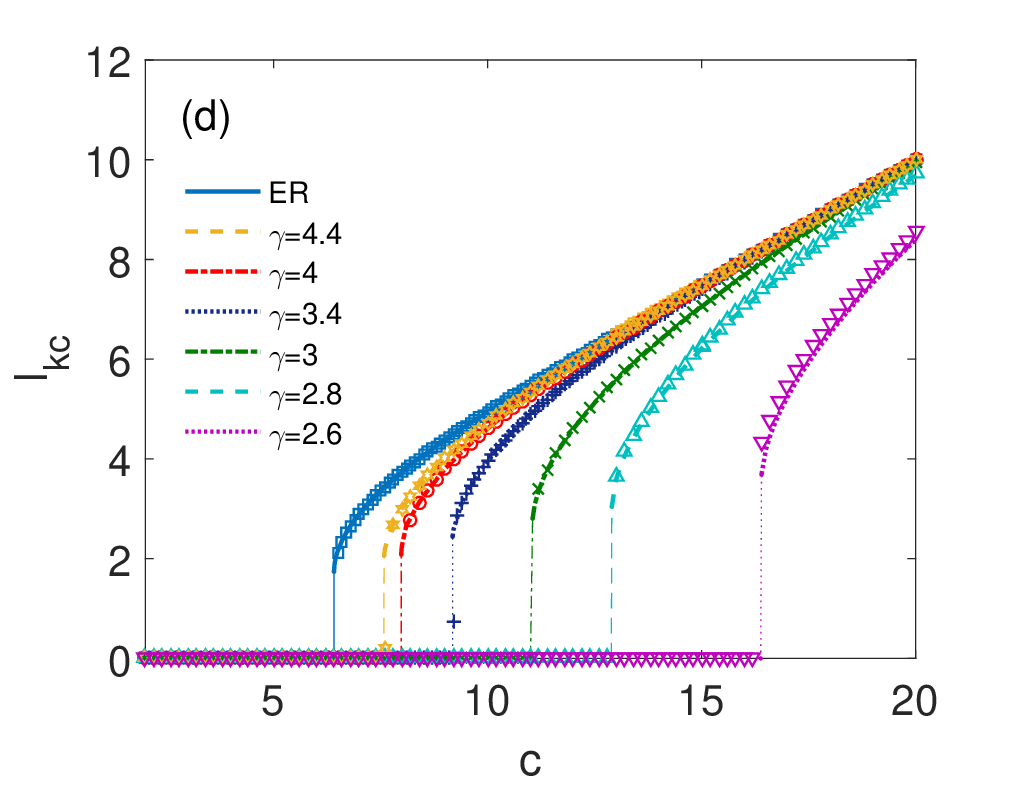}}
\scalebox{0.32}{\includegraphics[angle=0]{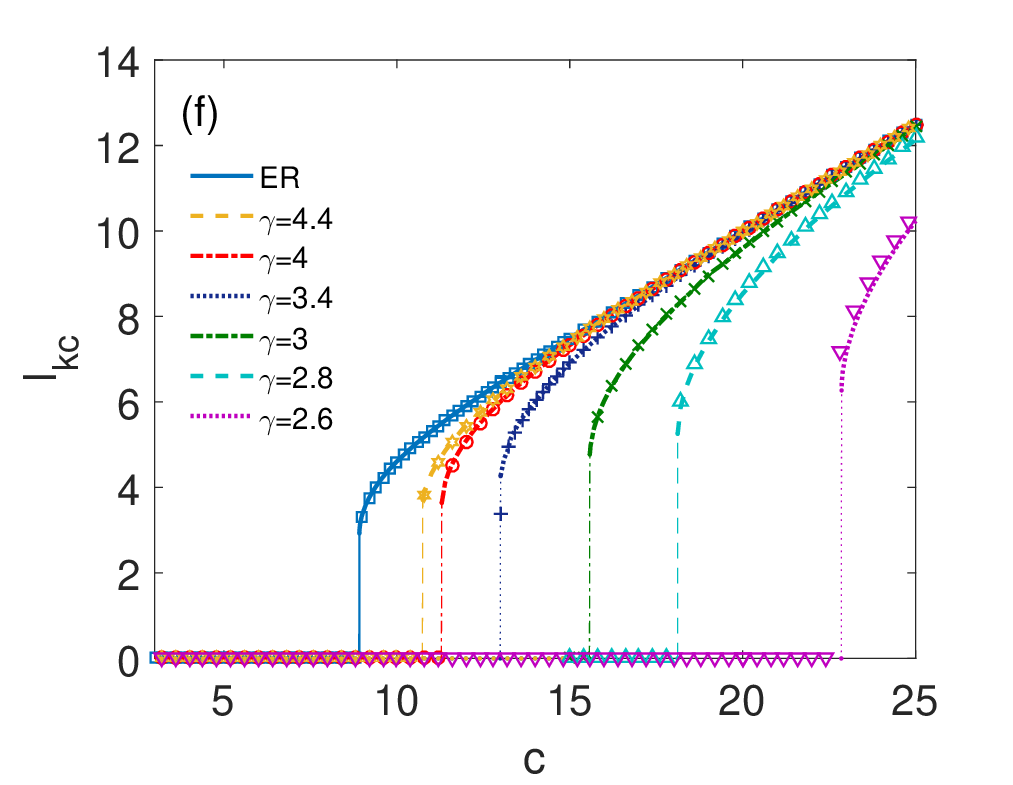}}
\end{center}
\caption{ (color online).
The relative sizes and the normalized number of edges of the $Gk$-core for $k=2,3,4$.
The points are the results of numerical simulation for the Erd\H{o}s--R\'enyi and asymptotically scale-free networks of size $N=10^{6}$, averaged over 10 realizations. The lines are analytical results obtained from Eqs.~(\ref{eq5})--(\ref{eq6}). As $\gamma$ approaches to 2, finite size effects become more important and a deviation between theoretical results and simulations is observed.
}
\label{f4}
\end{figure*}

Furthermore, the probability that both end vertices of an edge in the network belong to the $Gk$-core is $(1-\alpha-\beta)^2$. Hence, the fraction of edges in the $Gk-$core, denoted by $l_{kc}$, is obtained as
\begin{eqnarray}
l_{kc}&=&\frac{c}{2}(1-\alpha-\beta)^2.
\label{eq6}
\end{eqnarray}

Let us first consider Erd\H{o}s--R\'enyi networks with Poisson degree distributions, $P(q)=c^{q}e^{-c}/q!$, where $c$ is the vertex mean degree for the network. For the Poisson distribution, the generating function and its $s$-th derivative are $G(x)= e^{-c (1-x)}$ and $G^s(x)= c^{s}e^{-c(1-x)}$, respectively. One can easily find the relation between $\alpha$ and $\beta$ probabilities as $\beta= 1-e^{-c\alpha}$, which is independent of the value of $k$.
For Erd\H{o}s--R\'enyi networks with Poisson degree distributions, one can write a closed form for $\alpha$ and $n_{kc}$ from Eqs.~(\ref{eq2}) and (\ref{eq3}),
\begin{eqnarray}
\alpha&=&e^{-c\alpha}\, \frac{\Gamma[k-1,c(e^{-c\alpha} {-} \alpha)]}{(k-2)!}
,
\nonumber
\\[5pt]
n_{kc}&=&e^{-c\alpha} \Bigl\{ 1 - \frac{\Gamma[k,c(e^{-c\alpha} {-} \alpha)]}{(k-1)!}  \Bigr\}
.
\label{10}
\end{eqnarray}
In which, $\Gamma(s,x)$ is the upper incomplete gamma function.
\begin{figure}
\begin{center}
\scalebox{0.36}{\includegraphics[angle=0]{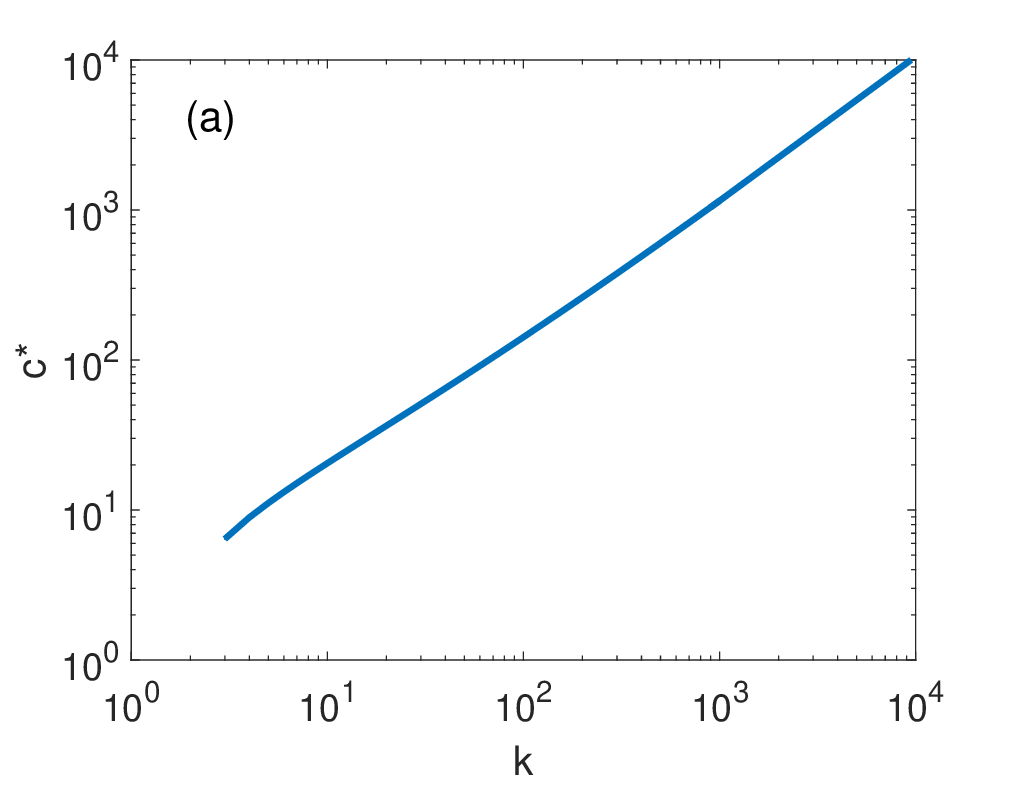}}~~~~\\
\scalebox{0.36}{\includegraphics[angle=0]{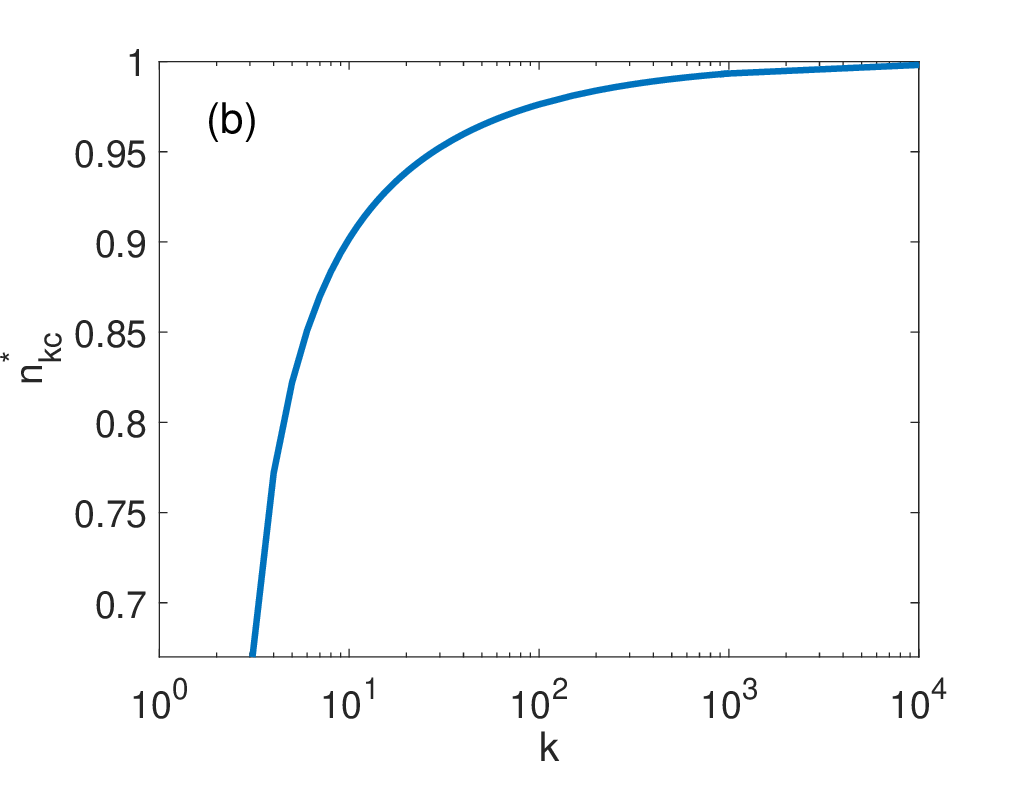}}~~~~
\end{center}
\caption{(color online). The behavior of $(a)$ the transition point $c^{*}$ and $(b)$ the size of the $Gk$-core at the transition point $n^{*}_{kc}$ vs. k for ER random networks.
}
\label{f33}
\end{figure}
The relative size and the normalized number of edges of the $Gk$-core for $k=2,3$ and $4$ are shown in Fig.~\ref{f4}. The analytic results (curves) are compared with numerical simulations (symbols). As we can see in the figure, in contrast to the ordinary core $(k=2)$, for $k\geq 3$ a $Gk$-core emerges discontinuously at the percolation threshold.

Equations ~(\ref{10}) enable us to obtain the transition point $c^{*}$ and the size of the $Gk$-core at the transition point, $n^{*}_{kc}$, for each $k$. From numerical data, we estimate the asymptotic representations for $c^{*}$ and $n^{*}_{kc}$ as the following:
\begin{eqnarray}
&&
c^*\approx k + C \sqrt{k} \ln\ln k
,
\nonumber
\\[3pt]
&&
n^{*}_{kc}\approx 1 - \frac{1}{C \sqrt{k} \ln\ln k}
,
\nonumber
\\[3pt]
&&
C=2.413...
.
\label{60}
\end{eqnarray}
\begin{figure*}[t]
\begin{center}
\scalebox{0.32}{\includegraphics[angle=0]{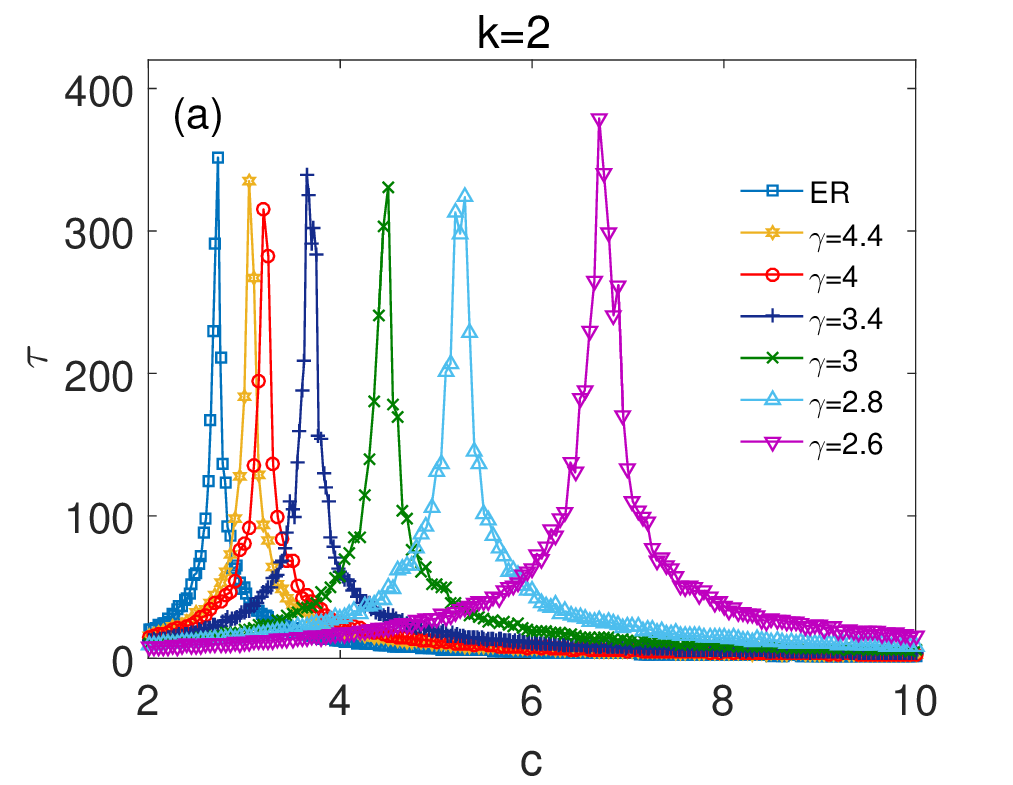}}
\scalebox{0.32}{\includegraphics[angle=0]{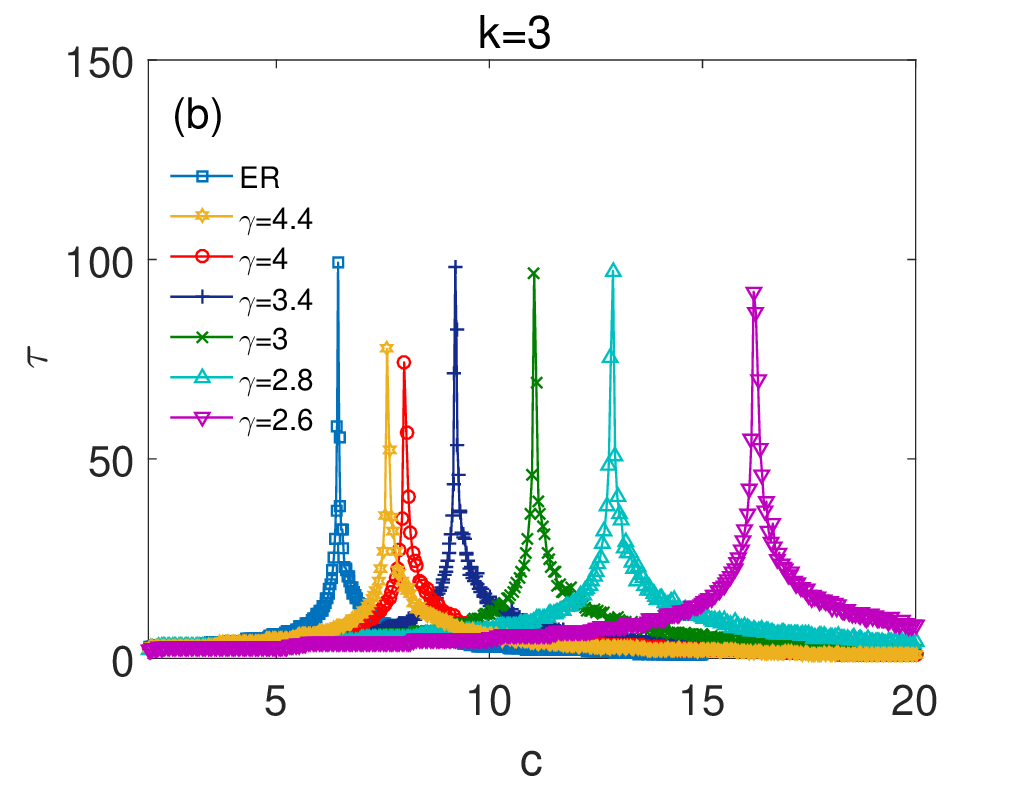}}
\scalebox{0.32}{\includegraphics[angle=0]{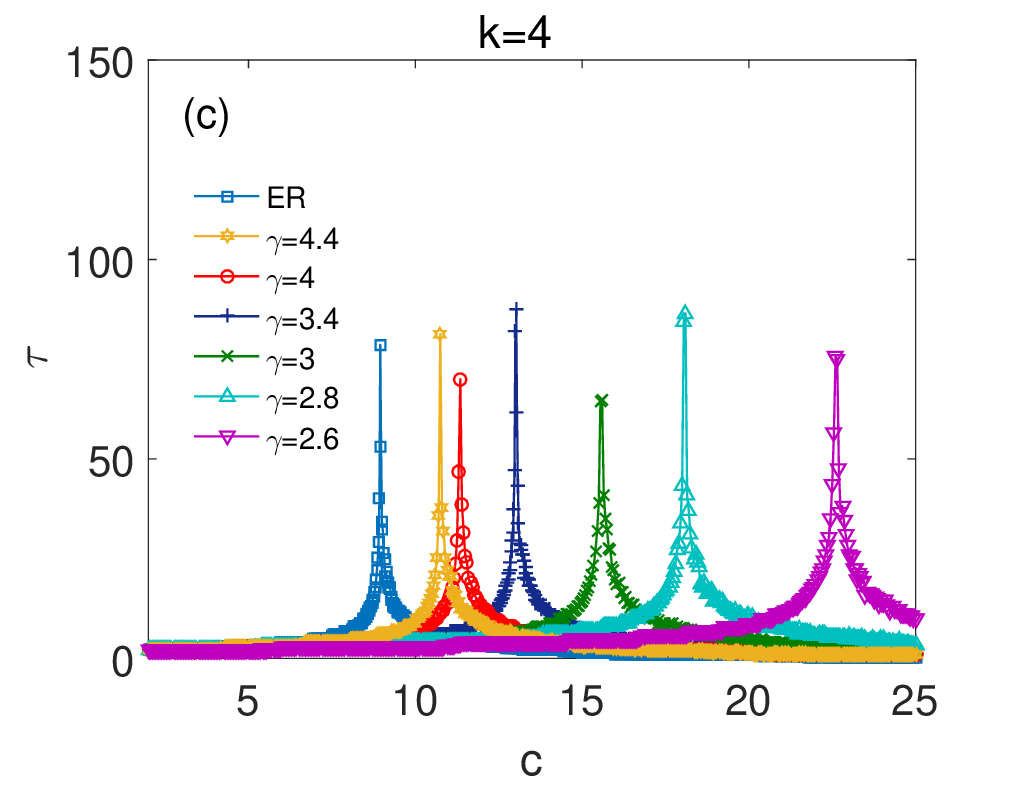}}
\end{center}
\caption{ (color online). Total number of pruning steps $\tau$ vs. mean degree $c$. The curves shows diverging of $\tau(c)$ at the emergence point of $(a)$ $G2$-core, $(b)$ $G3$-core and $(c)$ $G4$-core for the Erd\H{o}s--R\'enyi and asymptotically scale-free networks of size $N=10^{6}$, averaged over 10 realizations.
}
\label{f5}
\end{figure*}

Figure~\ref{f33} shows the behavior of $c^{*}$ and $n^{*}_{kc}$, in which the curves asymptotically coincide to Eqs.~(\ref{60}).

Next we consider scale-free networks. It was shown that for the purely power-law scale-free networks the ordinary core does not exist \cite{corepercolation}. Hence we consider the asymptotically scale-free, uncorrelated networks generated by the static model with the degree distribution
$P(q) =[\frac{c(\gamma-2)}{2(\gamma-1)}]^{\gamma-1}\Gamma(q-\gamma+1,\frac{c (\gamma-2)}{2(\gamma-1)})/\Gamma(q+1) \cong q^{-\gamma}$,
where $\Gamma(s)$ is the gamma function \cite{static, Goh:gkk01}. For this degree distribution the generating function is $G(x)=(\gamma-1)E_n[(1-x)\frac{c (\gamma-2)}{2(\gamma-1)}]$, where $E_n(x)=\int_{1}^{\infty}dy e^{-xy} y^{-n}$ is the exponential integral. Figure~\ref{f4} shows the relative size and the normalized number of the general $2$, $3$ and $4$-cores for different values of $\gamma$ versus c. With decreasing $\gamma$, the emergence point is shifted towards higher values of $c$. For scale free networks when $\gamma\rightarrow 2$, finite size effects become more significant. By imposing the proper degree cutoffs, one can eliminate the finite size effects and the intrinsic degree correlations \cite{cutoff1, cutoff2}. In Fig.~\ref{f4} we compare the emergence of cores
for asymptotically scale-free and Erd\H{o}s--R\'enyi networks. As one can see, the
dependence of the cores on $c$ for these networks is similar and,
as expected, the curves with larger $\gamma$ approach the result for
Erd\H{o}s--R\'enyi networks.

We define pruning time steps in a way that enables us to classify the vertices of the network into a set of layers for a given $k$. At time step $ t'=1$, we select the vertices of degree less than $k$ ($k$-leaves) and remove these vertices and their neighbors by applying the $k$-leaf algorithm. Removing the vertices in the first step may produce new $k$-leaves, which will be removed at $t'=2$ and so on. The vertices removed at each step $t'$ form a layer of the network. In other words, the network is pruned layer by layer until there is no $k$-leaf left. We denote the total number of pruning steps as $\tau$ so that $ t'=\{ 1,2, \ldots \tau\}$. After $\tau$ steps, the network consists of finite components or a giant $Gk$-core. For different networks we obtain $\tau(c)$ using numerical simulation, see in Fig.~\ref{f5}. As we can see, the dependencies $\tau(c)$ diverge at the birth points of the cores.

\section{Rate equations}
\label{rate}

The structural evolution of the network during pruning processes is described by the so-called rate equations for the degree distribution of the remaining network \cite{Weight:whbook2005,Weigt}. Here we derive rate equations for the $k$-leaf removal algorithm. Let us consider a network of $N$ vertices and $L$ edges. For simplicity we remove only the edges during the pruning process. In other words, at each time step $t$ we choose randomly a $k$-leaf, remove all $k$ edges incident to it, together with all edges incident to its $k$ neighbors. In this way, the number of vertices of the network remain constant. Note that the time steps $t$ differ from $t'$. The algorithm is iterated until $P(q)=0$ for all $q<k$. The important point of this approach is that the dynamics is self-averaging in the thermodynamic limit, i.e. $N\rightarrow \infty$. After a certain number of time steps, almost all random networks have the same degree distribution, which is independent of the (random) order of the removal of leaves \cite{Bauer}. Hence this approach can be used as a confirmation of the results obtained in the previous section.

We introduce the rescaled time $t=\frac{T}{N}$, where $T$ is the total number of steps of the pruning algorithm. So, $\Delta t = 1/N$ is the rescaled time of one iteration. Let $N(q,t)$ be the average number of vertices with degree $q$ at time $t$.  Since the total number of vertices is constant, i.e. $N(t)=N$, we have $N(q,t)=NP(q,t)$. We can write the change of $N(q,t+\Delta t)-N(q,t)$ after one iteration. In the large network limit, we can pass from the discrete difference to the time derivative of the degree distribution and obtain the following evolution equation for the degree distribution:
\begin{widetext}
\begin{eqnarray}
&&
 N(q,t+\Delta t)-N(q,t) =
\dot{P}(q,t)
\nonumber
\\[5pt]
&&
= - \frac{\theta(k-q)P(q,t)}{\sum_{q}\theta(k-q)P(q,t)}
+\delta _{q, 0}\Bigg(1+\frac{\sum_{q}q \theta(k-q)P(q,t)}{\sum_{q}\theta(k-q)P(q,t)}\Bigg)-\frac{\sum_{q}q \theta(k-q)P(q,t)}{\sum_{q}\theta(k-q)P(q,t)} \times \frac{qP(q,t)}{\langle q \rangle_{t}}
\nonumber
\\[5pt]
&&
+\frac{\sum_{q}q \theta(k-q)P(q,t)}{\sum_{q}\theta(k-q)P(q,t)}\times\frac{\sum_q q(q-1)P(q,t)}{\sum_q qP(q,t)}
\Bigg(\frac{(q+1)P(q+1,t)-qP(q,t)}{\langle q\rangle_t}\Bigg)
\label{master}
\end{eqnarray}
\end{widetext}

Let us explain different terms on the right hand side of Eq.~(\ref{master}). First we choose a random vertex of degree less than $k$ and remove all edges incident to it. The probability that a vertex has degree less than $k$ is $\frac{\theta(k-q)P(q,t)}{\sum_{q}\theta(k-q)P(q,t)}$, where $\theta(i)$ is defined for integers: $\theta(i{\geq}0)=1$ and $\theta(i<0)=0$. Thus with this probability, the number of vertices with $q<k$ decreases by 1. This gives the first term. After removing the edges incident to the leaf and all edges incident to its neighbors, the leaf and all its neighbors become vertices of degree zero. The average number of neighbors of a vertex of degree less than $k$ is $\frac{\sum_{q}q \theta(k-q)P(q,t)}{\sum_{q}\theta(k-q)P(q,t)}$. Hence the second term shows the number of vertices whose degrees become zero. On the other hand, the degree distribution of the end vertices of a randomly
chosen edge is $\frac{qP(q)}{\langle q\rangle}$. When we remove
the edges incident to the nearest neighbors of the leaf, the number of vertices of degree $q$ is decreased by the mean degree of the leaf with probability $\frac{qP(q)}{\langle q\rangle}$. Finally the last contribution is resulted from modification of degrees of the second neighbors of the leaf. After removal of all edges incident to the leaf and its nearest neighbors, the number of connections of the second nearest neighbors
of the leaf decreases by one. The average number of the second neighbors is equal to the mean degree of the nearest neighbors except one (connection to the leaf), multiplied by the average number of the nearest neighbors of the leaf.
Equation (\ref{master}) is a set of differential equations, describing
the evolution of a network during the pruning. For $k=2$, these equations coincide with the known ones \cite{Weigt}. Solving Eq.~(\ref{master}) iteratively, we can obtain the degree distribution of the network at each time step $t$.

As we already mentioned, we do not remove the vertices during the leaf removal algorithm and so the total number of the vertices remains constant. However, at each time step all edges incident to the leaf and also the edges incident to all its nearest neighbors are removed. Hence, at each time step the average number of removed edges is equal to the mean number of nearest neighbors multiplied by their mean degree. This results to the following evolution equation for the average number of remained edges in the network:
\begin{eqnarray}
\frac{\dot{L}(t)}{N}&=&-\frac{\langle q^{2}\rangle_{t}}{\langle q\rangle_{t}}\times \frac{\sum_{q}q \theta(k-q)P(q,t)}{\sum_{q}\theta(k-q)P(q,t)}
\label{edges}
\end{eqnarray}

We apply the leaf removal algorithm to an uncorrelated network with a degree distribution $P(q,t = 0)$ and
a vertex mean degree equal to $c_0$ as the initial conditions. For each value of $k$, the
algorithms are iterated until no vertices of degree less than $k$ remain. To find the $Gk$-core, the algorithm must continue until time $t^{*}_k$ at which $P(1,t_k^*)=P(2,t_{k}^*)=\ldots=P(k-1,t^{*}_{k})=0$. Our numerical results for different networks show that $P(1,t)$ is the last probability to become zero. That is, the vertices of degree $1$ disappear after all other leaves. This is why during iteration we look at the behavior of $P(1,t)$ and the algorithm stops at time $t^{*}_{k}$  for a given $k$. The remaining subgraph is the $Gk$-core. For $k=2$ the algorithm coincides with the ordinary leaf-removal algorithm and the remaining subgraph is the $G2$-core or simply the core. After we find $t_k^*$, the size
and the number of edges of the $Gk$-core can be obtained from the following relations:
\begin{eqnarray}
N_{kc}&=& N[1-P(0,t_k^{*})]
,
\label{eq11}
\\[5pt]
L_{kc}&=&L(t_k^{*})
.
\label{eq12}
\end{eqnarray}
\begin{figure}[t]
\begin{center}
\scalebox{0.38}{\includegraphics[angle=0]{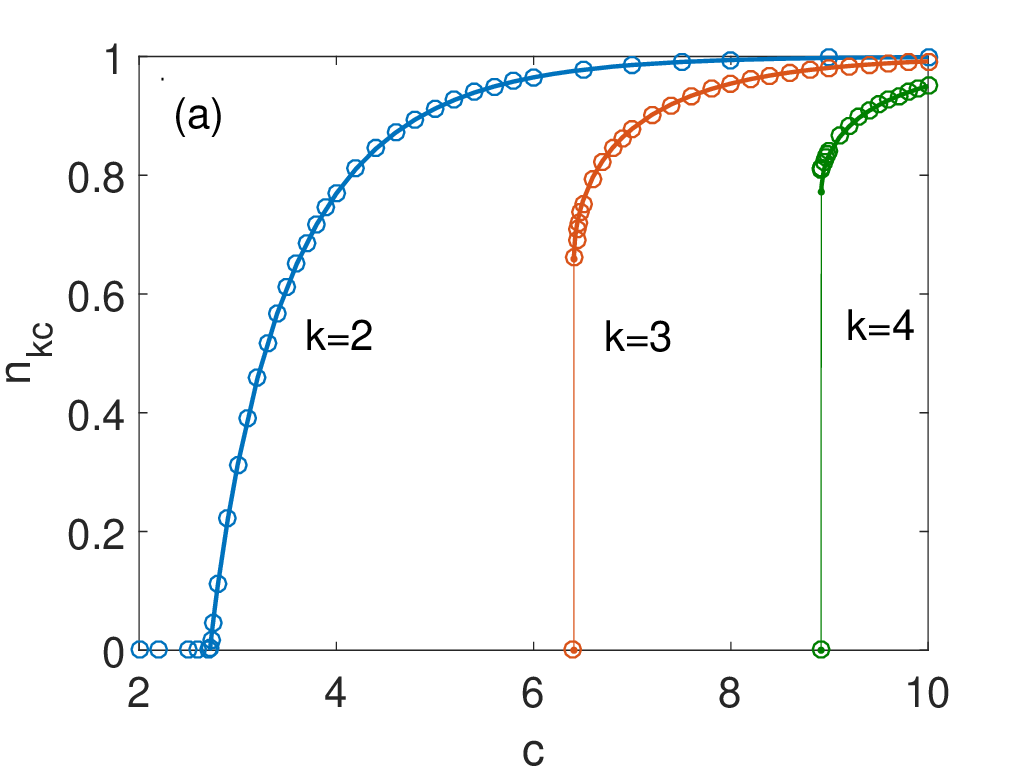}}
\scalebox{0.38}{\includegraphics[angle=0]{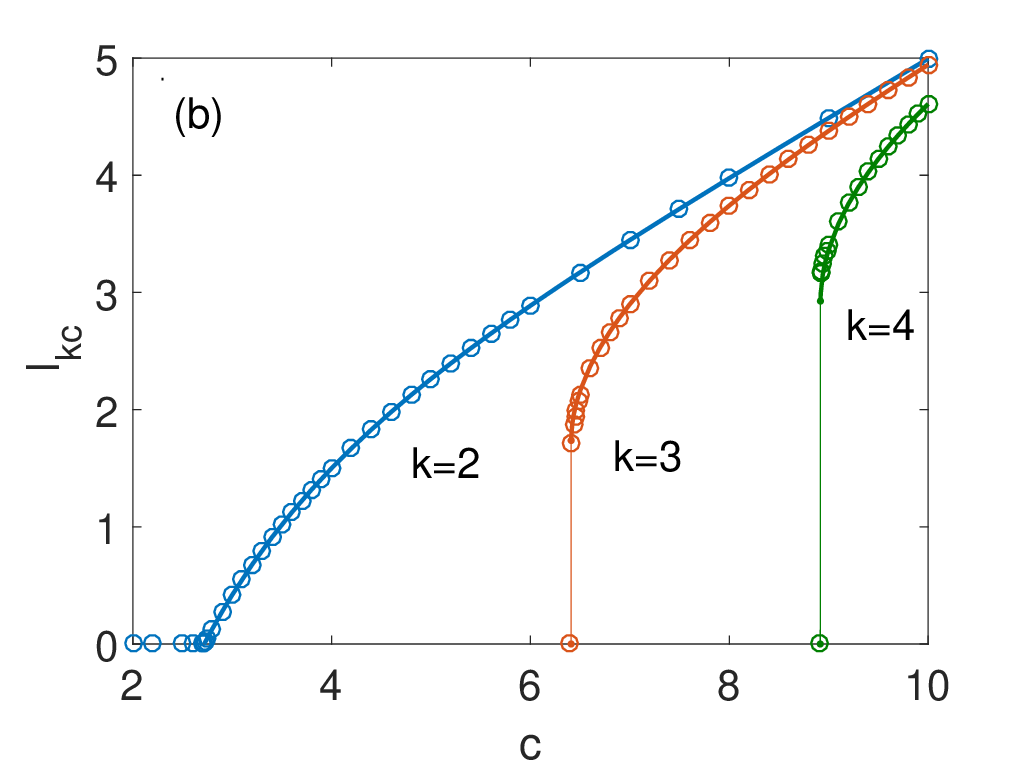}}
\end{center}
\caption{(color online). $(a)$ The relative sizes and $(b)$ normalized number of edges of the $Gk$-core in Erd\H{o}s--R\'enyi networks for $k=2,3$ and $4$, vs the mean degree $c$ of the network. The points show the results obtained by the rate equation approach and lines show the results obtained using the formalism of Sec.~II (Eqs.~(\ref{eq5})--(\ref{eq6})).
}
\label{f6}
\end{figure}

We apply this approach to the Erd\H{o}s--R\'enyi random graphs. The Poisson degree
distribution rapidly decays and it is sufficient to consider $q_{max} = 30$, i.e. we solve the set of the first $31$ equations. Figure~\ref{f6} shows the size and the number
of edges calculated from Eqs.~(\ref{master})--(\ref{eq12}), for the Erd\H{o}s--R\'enyi networks. In this figure we compare the results obtained by solving the rate equations (points) with the analytic results of the previous section (lines) for the general $2$, $3$ and $4$-cores.
\begin{table*}[t]\centering
\begin{tabular}{ |p{3cm}||p{1cm}|p{1cm}|p{0.7cm}|p{0.8cm}|p{1.5cm}|p{1.5cm}|p{1.3cm}|p{1.3cm}|  }
 \hline
 Name& N &L& Ref&  $k_{max}$& $n_{k_{max}-core}$&$l_{k_{max}-core}$&$n_{2-core}$&$l_{2-core}$\\
 \hline
 E. Coli, transcription  & 97 &212& \cite{coli}&3 &0.319 &0.793 &0.917 &2.051  \\
 AS Oregon& 6474  & 12572 &\cite{as} & 2&0.001 &0.001 &0.001 &0.001 \\
 Astrophysics &16046 & 121251& \cite{condmat} & 31&0.002 & 0.045& 0.769& 5.980\\
 C. Elegans, neural &297 & 2148&\cite{elegan} &3&0.885 &6.037 &0.915 &6.447\\
 Cond-Mat& 16264  & 47594& \cite{condmat}& 10&0.006 &0.003 &0.618 &1.884\\
 Dolphins& 62  & 159 &\cite{dolphin} &3& 0.161& 0.290& 0.645& 1.322 \\
 Email-Enron& 36692 & 183831& \cite{email} & 7&0.0004 &0.001 &0.389 &1.052\\
 Linux& 30834  & 213217&\cite{linux} & 5&0.0003 &0.0008 &0.147 &0.375\\
 petster-friendship-hamster& 1858  & 12534&\cite{linux} & 8& 0.010&0.047 &0.584 &2.664\\
 Sociopatterns-Infectious& 410  &2765 &\cite{socio} &9 &0.056 &0.443 &0.912 &6.090 \\
 PGPgiantcompo& 10680  & 24316& \cite{PGP}&17 &0.001 &0.014 &0.158 & 0.483\\
 US Air Trasportation& 500  & 2980& \cite{metapopulation}&3 &0.008 &0.012 & 0.260& 0.494\\
 Yeast- protein & 2284 &6646&\cite{yeast} & 3&0.003 &0.011 &0.025 & 0.052\\
 \hline
\end{tabular}
\caption{$Gk$-core decomposition of real networks with the number of vertices $N$ and the number of edges $L$. $k_{max}$ is the label of the innermost core. $n_{k_{max}-core}$ and $n_{2-core}$ show the relative size of the innermost and outermost cores respectively. Similarly, $l_{k_{max}-core}$ and $l_{2-core}$ show the normalized number of edges of the innermost and outermost cores respectively. }
\label{t1}
\end{table*}
\begin{figure*}[t]
\begin{center}
\scalebox{0.4}{\includegraphics[angle=0]{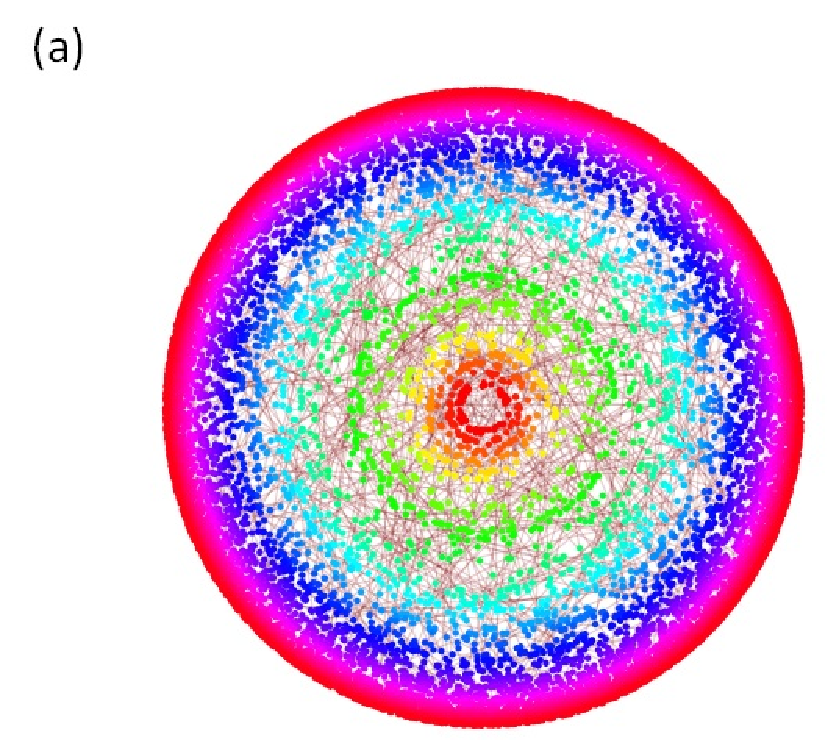}}~~~~~~~~~~~~~
\scalebox{0.38}{\includegraphics[angle=0]{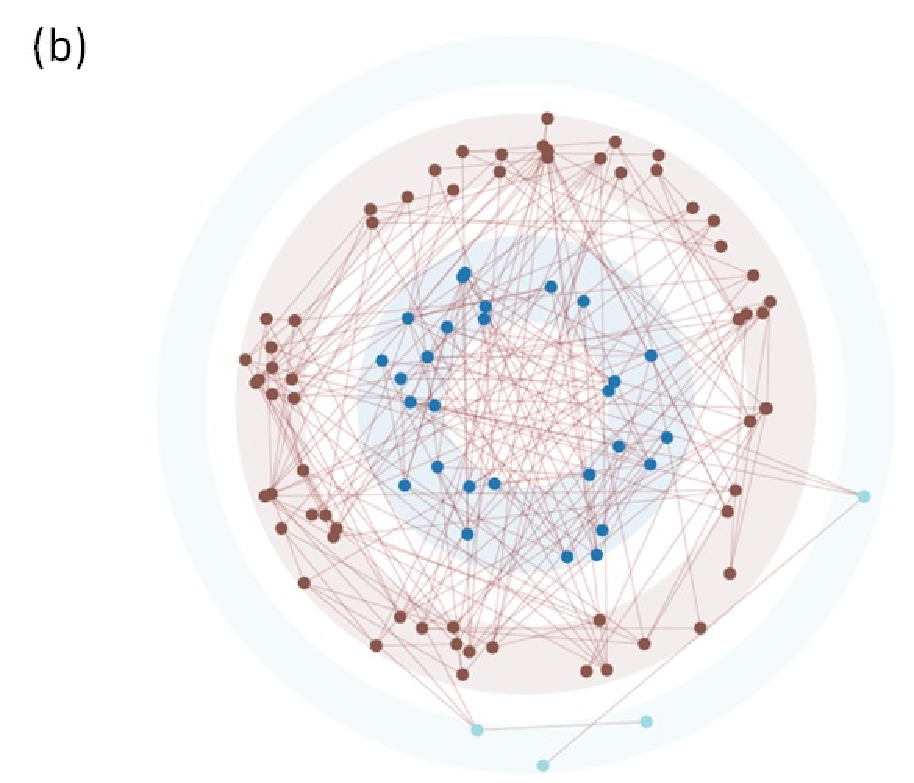}}
\end{center}
\caption{(color online). Graphical visualization of the $Gk$-core decomposition of $(a)$ astrophysics and $(b)$ transcriptional regulation networks.
}
\label{f8}
\end{figure*}
\section{real-world networks}
\label{real}

We apply the $k$-leaf removal to a number of real-world networks and find cores of these networks. The characteristics of real-world networks, analyzed in the paper, are listed in Table \ref{t1}.
The outermost core is the largest core which corresponds to $k=2$ and includes other cores. As we increase the value of $k$, the size of cores is decreased and the core corresponding to maximum $k$ ($k_{max}$) is the smallest and innermost core. We present the relative size and number of edges of the outermost and innermost $Gk$-cores
 in Table \ref{t1}.
We find that many real social networks are decomposed to a large hierarchy of the $Gk$-cores. For instance the layers of arxiv networks, i.e. cond-mat, astro-ph, hep-th, $\ldots$, have the highest numbers of the $Gk$-cores nested into each other among networks analyzed in this paper. In contrast, the food webs and biological networks have the a small number of cores. Using the visualization algorithm proposed in \cite{Alvarez}, visualization of astrophysics network in 2005 \cite{condmat} and transcriptional regulation network \cite{coli} are presented as two examples in Fig.~\ref{f8}. The regulation network has a few cores while the astrophysics network has around 30 cores in our proposed network decomposition scheme. Comparing with the random networks, the real networks have more cores. Similar to what was found in the ordinary core problem, this difference reveals that other structural features such as correlations and clustering may be significant for the sizes and organization of the $Gk$-cores.
\section{Conclusion}
In this work we have generalized the ordinary core subgraph to the $Gk$-cores. We proposed the $k$-leaf removal algorithm as a generalization of the ordinary leaf removal
The $k$-leaf pruning algorithm enables us to decompose large random networks into a hierarchical set of progressively nested subgraphs which we called the $Gk$-cores. Our approach can also be considered as a generalization of the ordinary $k$-core decomposition. In our pruning at each time step, not only the vertices of degree less than $k$ but also their nearest neighbors are removed. Following this pruning, the network is decomposed into a hierarchy of progressively nested $Gk$-cores such that the vertices, belonging to the inner cores, and also their first neighbors are of higher degree and well connected. Using generating function technique, we found the structural characteristics and the emergence point of the $Gk$-cores for the Erd\H{o}s--R\'enyi and scale-free random networks. Similarly to the ordinary $k$-core percolation, $Gk$-cores show a discontinuous phase transition for $k\geq 3$. We compared our results with numerical simulations and observed
a complete agreement. In addition, we used the rate equation approach to describe the evolution of degree distribution of random networks during the $k$-leaf pruning algorithm. We checked that the result of the application of this approach to the Erd\H{o}s--R\'enyi graph completely coincides with the exact result obtained by the analytical calculations.
We have applied the $k$-leaf removal algorithm to a number of real-world networks. Among the real networks explored, the social networks have a large $k_{max}$.

We emphasize that in contrast to the $k$-core decomposition, the $Gk$-cores are not about the classification of vertices in a network according to their properties but rather about the characterization of a specific robustness of this network. Suppose that a network be attacked by a virus infecting/removing weak vertices (of degree less than $k$) and their nearest neighbors. The $Gk$-cores show what will remain of the network after this epidemics. The resilience/robustness of a network against this kind of epidemics is characterized by the size of its $Gk$-core.  This may explain why the social networks that we explored have a large $k_{max}$.


\begin{thebibliography}{99}

\bibitem{Alvarez}
I. Alvarez-Hamelin, L. Dell-Asta, A. Barrat, and A. Vespignani, {\em K-core Decomposition of Internet Graphs: Hierarchies, Self-Similarity and Measurement Biases}, arXiv.cs.NI/0511035 (2005).

\bibitem{Carmi}
S. Carmi, S. Havlin, S. Kirkpatrick, Y. Shavitt, and E. Shir, {\em  A model of Internet topology
using k-shell decomposition}, Proc. Natl. Acad. Sci. USA {\bf 104}, 11150 (2007).

\bibitem{Seidman}
S.B. Seidman, {\em Network Structure and Minimum Degree}, Social Networks {\bf 5}, 269 (1983).

\bibitem{Kitsak}
M. Kitsak, L. K. Gallos, S. Havlin, F. Liljeros, L. Muchnik,
H. E. Stanley, and H. A. Makse, {\em Identification of Influential Spreaders in Complex Networks}, Nature Phys. {\bf 6}, 888 (2010).

\bibitem{k-corepercolation}
S. N. Dorogovtsev, A. V. Goltsev, and J. F. F. Mendes, {\em K-core Organization of Complex Networks} ,Phys.
Rev. Lett. {\bf 96}, 040601 (2006).

\bibitem{critical}
S. N. Dorogovtsev, A. V. Goltsev, and J. F. F. Mendes, {\em Critical Phenomena in Complex Networks},
Rev. Mod. Phys. {\bf 80}, 1275 (2008).

\bibitem{generalk-core1}
N. K. Panduranga, J. Gao, X. Yuan, H. E. Stanley, and S. Havlin, {\em Generalized Model for
k-core Percolation and Interdependent Networks}, Phys. Rev. E. {\bf 96}, 032317 (2017).

\bibitem{generalk-core2}
N. Azimi-Tafreshi, J. G\'omez-Garde\~nes, and S. N. Dorogovtsev, {\em k-core Percolation on Multiplex Networks},
Phys. Rev. E. {\bf 90}, 032816 (2014).

\bibitem{Karp}
R. M. Karp and M. Sipser, {\em Maximum Matching in Sparse Random Graphs},
Proc. 22nd Annual IEEE Symp. on Foundations of Computer Science pp.
364-375 (1981).

\bibitem{Bauer}
M. Bauer and O. Golinelli, {\em Core Percolation in Random Graphs: A Critical Phenomena Analysis},
Eur. Phys. J. B {\bf 24}, 339
(2001).

\bibitem{local}
M. Bauer and O. Golinelli, {\em Exactly Solvable Model with Two Conductor-Insulator Transitions Driven by Impurities},
Phys. Rev. Lett. {\bf 86}, 2621
(2001).


\bibitem{replica}
M. Weigt and A.K. Hartmann, {\em Number of Guards Needed by a Museum: a Phase Transition in Vertex Covering of Random Graphs},
Phys. Rev. Lett. {\bf 84} 6118 (2000).

\bibitem{control1}
Y. -Y. Liu,  J.-J. Slotine, and A.-L. Barab\'asi, {\em Controllability of complex networks},
Nature {\bf 473}, 167 (2011).

\bibitem{control2}
T. Jia, and M. P\'osfai, {\em Connecting Core Percolation and Controllability of Complex Networks},
Sci. Rep. {\bf4}, 5379 (2014).


\bibitem{Weight:whbook2005}
M. Weigt and A. K. Hartmann,
{\em Phase Transitions in Combinatorial Optimization Problems},
(Wiley-VCH, Weinheim, 2005).


\bibitem{MDS}
J. H. Zhao, Y. Habibulla, and H. J. Zhou, {\em Statistical Mechanics of the Minimum Dominating
Set Problem}, J. Stat. Phys. {\bf 159}, 1154 (2015).

\bibitem{Zhao}
J. H. Zhao, H. J. Zhou and Y. Y. Liu, {\em Inducing Effect on the Percolation Transition in Complex Networks}, Nat. Commun. {\bf 4}, 2412 (2013).




\bibitem{Weigt}
M. Weigt, {\em Dynamics of Heuristic Optimization Algorithms on Random Grapghs},
Eur. Phys. J. B {\bf 28}, 369 (2002).


\bibitem{generating}
M. E. J. Newman, S. H. Strogatz, and D. J. Watts, {\em Random Graphs with Arbitrary Degree Distributions and Their Applications},
Phys. Rev. E {\bf 64}, 026118 (2001).


\bibitem{coredir}
N. Azimi-Tafreshi, S. N. Dorogovtsev, and J. F. F. Mendes, {em Core Organization of Directed Complex Networks},
Phys. Rev. E. {\bf 87}, 032815 (2013).


\bibitem{corepercolation}
Y.-Y. Liu, E. Cs\'oka, H. Zhou, and M. P\'osfai, {\em Core Percolation on Complex Networks},
Phys. Rev. Lett. {\bf 109}, 205703 (2012).


\bibitem{static}
M. Catanzaro and R. Pastor-Satorras, {\em Analytic Solution of a Static Scale-Free Network Model},
Eur. Phys. J. B {\bf 44}, 241 (2005).


\bibitem{Goh:gkk01}
K. -I. Goh, B. Kahng, and D. Kim, {\em Universal Behavior of Load Distribution in Scale-Free Networks},
Phys. Rev. Lett. {\bf 87}, 278701 (2001).


\bibitem{cutoff1}
M. Bogu\~n\'a, R. Pastor-Satorras,and A. Vespignani, {\em Cut-offs and finite size effects in scale-free networks}, Eur. Phys. J. B {\bf 38}, 205 (2004).

\bibitem{cutoff2}
J. -S. Lee, K. -I. Goh, B. Kahng and D. Kim, {\em Intrinsic degree-correlations in the
static model of scale-free networks} Eur. Phys. J. B {\bf 49}, 231 (2006).



\bibitem{coli}
S. S. Shen-Orr, R. Milo, S. Mangan, and U. Alon, {\em Network motifs in the transcriptional regulation network of Escherichia coli} ,Nat. Genet. {\bf 31}, 64 pmid:11967538 (2002).


\bibitem{as}
J. Leskovec, J. Kleinberg, and C. Faloutsos, KDD '05, {\em Graphs over Time: Densification Laws, Shrinking Diameters and Possible Explanations}, Proceedings of the Eleventh ACM SIGKDD International Conference on Knowledge Discovery in Data Mining 177 (2005).


\bibitem{condmat}
M. E. J. Newman, {\em The Structure of Scientific Collaboration Networks}, Proceedings of the National Academy of Sciences {\bf 98}(2), 404 (2001).

\bibitem{elegan}
J. D. Watts, and S. H. Strogatz, {\em Collective Dynamics of Small-World Networks}, Nature {\bf 393} 440 (1998).

\bibitem{dolphin}
D. Lusseau, K. Schneider, O. J. Boisseau, P. Haase, E. Slooten, and S. M. Dawson, {\em The Bottlenose Dolphin Community of Doubtful Sound Features a Large Proportion of Long-Lasting Associations}, Behavioral Ecology and Sociobiology {\bf 54} (4), 396 (2003).


\bibitem{email}
M. W. Mahoney, A. Dasgupta, K. J. Lang, and J. Leskovec, {\em Community Structure in Large Networks: Natural Cluster Sizes and the Absence of Large Well-Defined Clusters}, Internet Mathematics {\bf 6} 29 (2009).


\bibitem{linux}
J. Kunegis, {\em KONECT: The Koblenz Network Collection}, WWW '13 Companion, Proceedings of the 22Nd International Conference on World Wide Web, 1343 (2013).


\bibitem{socio}
L. Isella, J. Stehl\'e, A. Barrat, C. Cattuto, J.-F. Pinton, and W. Van den Broeck, {\em What’s in a Crowd? Analysis of Face-to-Face Behavioral Networks}, J. Theor. Biol. {\bf271}, 166 (2011).

\bibitem{PGP}
M. Bogu\~n\'a, R.  Pastor-Satorras, A. Di\'az-Guilera and A. Arenas, {\em Models of Social Networks Based on Social Distance Attachment}, Phys. Rev. E {\bf 70}, 056122 (2004).

\bibitem{metapopulation}
V. Colizza, R. Pastor-Satorras and A. Vespignani, {\em Reaction-Diffusion Processes and Metapopulation Models in Heterogeneous Networks}, Nature Physics {\bf 3}, 276 (2007).


\bibitem{yeast}
D. Bu, Y. Zhao, L. Cai, H. Xue, X. Zhu, H. Lu, J. Zhang, S. Sun, L. Ling, N. Zhang, G. Li, and R. Chen1, {\em Topological Structure Analysis of the Protein-Protein Interaction Network in Budding Yeast}, Nucleic Acids Res. {\bf 31}(9) 2443 (2003).





\end{thebibliography}
\end{document}